\documentclass[aps,prb,twocolumn,superscriptaddress,oneside,floatfix,amsmath,showpacs,amssymb]{revtex4}
\usepackage{graphicx}
\usepackage{dcolumn}
\usepackage{bm}

\begin{document}

\newcommand{\PF}{(TMTSF)$_2$PF$_6$}
\newcommand{\ClO}{(TMTSF)$_2$ClO$_4$}
\newcommand{\X}{(TMTSF)$_2$X}
\newcommand{\ie}{\textit{i.e.}}
\newcommand{\eg}{\textit{e.g.}}
\newcommand{\etal}{\textit{et al.}}
\newcommand{\tsdw}{$T_{\textrm{SDW}}$}


\title{Correlation between linear resistivity and $T_c$ in the Bechgaard salts and the pnictide superconductor Ba(Fe$_{1-x}$Co$_x$)$_2$As$_2$}


\author{Nicolas~Doiron-Leyraud}
\email{ndl@physique.usherbrooke.ca}
\affiliation{D\'epartement de physique and RQMP, Universit\'e de Sherbrooke, Sherbrooke, Qu\'ebec, J1K 2R1 Canada}

\author{Pascale~Auban-Senzier}
\affiliation{Laboratoire de Physique des Solides, UMR 8502 CNRS Universit\'e Paris-Sud, 91405 Orsay, France}

\author{Samuel~Ren\'e~de~Cotret}
\affiliation{D\'epartement de physique and RQMP, Universit\'e de Sherbrooke, Sherbrooke, Qu\'ebec, J1K 2R1 Canada}

\author{Claude~Bourbonnais}
\affiliation{D\'epartement de physique and RQMP, Universit\'e de Sherbrooke, Sherbrooke, Qu\'ebec, J1K 2R1 Canada}
\affiliation{Canadian Institute for Advanced Research, Toronto, Canada}

\author{Denis~J\'erome}
\affiliation{Laboratoire de Physique des Solides, UMR 8502 CNRS Universit\'e Paris-Sud, 91405 Orsay, France}
\affiliation{Canadian Institute for Advanced Research, Toronto, Canada}

\author{Klaus~Bechgaard}
\affiliation{Department of Chemistry, H.C. {\O}rsted Institute, Copenhagen, Denmark}

\author{Louis~Taillefer}
\email{Louis.Taillefer@USherbrooke.ca}
\affiliation{D\'epartement de physique and RQMP, Universit\'e de Sherbrooke, Sherbrooke, Qu\'ebec, J1K 2R1 Canada}
\affiliation{Canadian Institute for Advanced Research, Toronto, Canada}

\date{\today}


\begin{abstract}

The quasi-1D organic Bechgaard salt \PF~displays spin-density-wave (SDW) order and superconductivity in close proximity in the temperature-pressure phase diagram. We have measured its normal-state electrical resistivity $\rho_a(T)$ as a function of temperature and pressure, in the $T \to 0$ limit. At the critical pressure where SDW order disappears, $\rho_a(T) \propto T$ down to the lowest measured temperature (0.1~K). With increasing pressure, $\rho_a(T)$ acquires a curvature that is well described by $\rho_a(T) = \rho_0 + AT + BT^2$, where the strength of the linear term, measured by the $A$ coefficient, is found to scale with the superconducting transition temperature $T_c$. This correlation between $A$ and $T_c$ strongly suggests that scattering and pairing in \PF~have a common origin, most likely rooted in the antiferromagnetic spin fluctuations associated with SDW order. Analysis of published resistivity data on the iron-pnictide superconductor Ba(Fe$_{1-x}$Co$_x$)$_2$As$_2$ reveals a detailed similarity with \PF, suggesting that antiferromagnetic fluctuations play a similar role in the pnictides.

\end{abstract}

\maketitle


A number of strongly correlated metals share a common property: their electrical resistivity grows linearly with temperature $T$ from $T = 0$, in stark contrast with the standard Fermi liquid description of metals. Notorious examples of materials showing this linear resistivity are the high-$T_c$ cuprate superconductors~\cite{Takagi,Manako,Ando, Ito}, such as hole-doped La$_{1.6-x}$Nd$_{0.4-x}$Sr$_x$CuO$_4$ (Nd-LSCO)~\cite{Daou} and electron-doped Pr$_{2-x}$Ce$_x$CuO$_4$~\cite{Fournier}, near their ``stripe''~\cite{Ichikawa} and antiferromagnetic quantum critical point, respectively, and a number of quantum-critical heavy-fermion metals~\cite{vL}, such as CeCu$_{6-x}$Au$_x$~\cite{vL}, CeCoIn$_5$~\cite{Tanatar}, and YbRh$_2$Si$_2$~\cite{Trovarelli}. But the origin of this phenomenon remains a subject of debate because it has not yet been observed in a material whose ground state is well understood, without the complication of a pseudogap phase, a nearby Mott insulator, or $f$-electron moments and the associated Kondo effect. On the other hand, theoretical efforts are faced with a major puzzle: while the scattering rate at antiferromagnetic hot spots is linear in temperature, it is not clear how it will affect the electrical resistivity since on the remaining segments of the Fermi surface it has the usual quadratic temperature dependence~\cite{Hlubina}. Thus, while beautifully simple in appearance, the linear resistivity of cuprates and heavy-fermion metals remains a major open question in the physics of correlated electrons.

\begin{figure}[t]
\begin{center}
\includegraphics[scale=1.1]{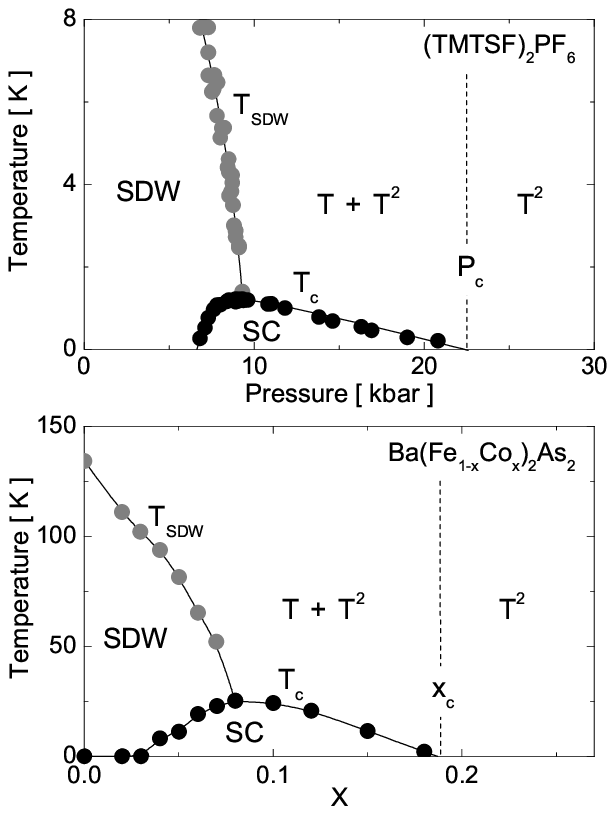}
\caption{Top: Temperature-pressure phase diagram of \PF, showing a spin-density-wave (SDW) phase below \tsdw~(grey dots) and superconductivity (SC) below $T_c$ (black dots) (from~\cite{Vuletic,Salameh} and this work). The latter phase ends at the critical pressure $P_c$. Bottom: Temperature-doping phase diagram of the iron-pnictide superconductor Ba(Fe$_{1-x}$Co$_x$)$_2$As$_2$ reproduced from \cite{Fang}, as a function of nominal Co concentration $x$, showing a metallic SDW phase below \tsdw~and superconductivity below a $T_c$ which ends at the critical doping $x_c$. In both panels the vertical dashed line separates a regime where the resistivity $\rho(T)$ grows as $T^2$ (on the right) from a regime where it grows as $T + T^2$ (on the left) (see text).
}
\label{fig1}
\end{center}
\end{figure}

We have examined this issue by studying the archetypal quasi-1D organic Bechgaard salt \PF~\cite{Jerome,Brown,Bourbonnais}, whose phase diagram is shown in Fig.~\ref{fig1}. The conducting chains of organic molecules give \PF~a strong quasi-1D character, reflected in its Fermi surface, made up of two slightly warped parallel sheets that nest well. As a result, \PF~orders in a spin-density-wave (SDW) state below a temperature \tsdw~$\approx$~12~K, which gets suppressed with pressure as next-nearest chain hopping is enhanced. As \tsdw~falls, superconductivity rises and peaks with $T_c$~$\approx$~1.2 K at the point where \tsdw~$\rightarrow$~0~\cite{Vuletic,Salameh}, forming a dome that extends to above 20~kbar. The phase diagram of \PF, with its adjacent semi-metallic SDW and superconducting phases, therefore closely resembles that of the iron-pnictide superconductor Ba(Fe$_{1-x}$Co$_x$)$_2$As$_2$ (see Fig. \ref{fig1}) and, to some extent, that of certain heavy-fermion metals \cite{vL,Mathur} and cuprates \cite{Daou,Ichikawa}. But a significant advantage of the Bechgaard salts is their relative simplicity. They are free from Kondo and Mott physics and, owing to their single quasi-1D Fermi surface, weak-coupling theory provides a good description of their electronic properties, in particular, the superconducting phase on the border of SDW order \cite{Bourbonnais,Sedeki}.


Here we report measurements of the $a$-axis electrical resistivity in \PF, \ie, along the chains of organic molecules, at low temperature as a function of pressure. Single crystals of \PF~were grown by the usual method of electrocrystallization~\cite{Bechgaard}. The samples used have typical values of $a$-axis conductivity near 500 ($\Omega$ cm)$^{-1}$ at room temperature and pressure. Typical sample dimensions are 1.5 x 0.2 x 0.05 mm$^3$ with the length, width and thickness along the $a$, $b$, and $c$ crystallographic axes, respectively. The current was applied along the $a$-axis and the magnetic field along the $c$-axis. Electrical contacts were made with evaporated gold pads (typical resistance between 1 and 10 $\Omega$) to which 17 $\mu$m gold wires were glued with silver paint. The electrical resistivity was measured with a resistance bridge using a standard four-terminal AC technique. Low excitation currents of typically 30 $\mu$A were applied in order to eliminate heating effects caused by the contact resistances. This was checked using different values of current above and below this value, at temperatures below 1 K. A non-magnetic piston-cylinder pressure cell was employed~\cite{Walker}, with Daphne oil 7373 as pressure transmitting medium. The pressure at room temperature and 4.2 K was measured using the change in resistance and superconducting $T_c$ of a Sn sample, respectively. Only the values recorded at 4.2 K are quoted here.


\begin{figure}[t!]
\begin{center}
\includegraphics[scale=1.2]{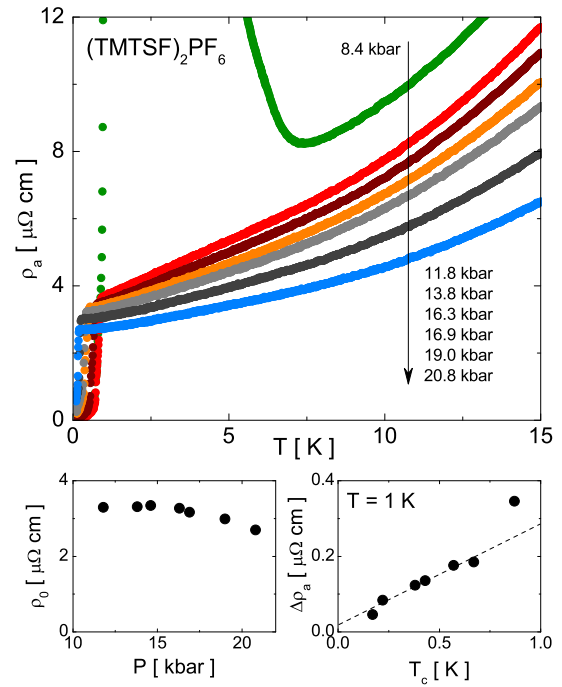}
\caption{(Color online) Top: $a$-axis electrical resistivity $\rho_a(T)$ of \PF~as a function of temperature at various pressures as indicated. Bottom left: residual resistivity $\rho_0$ as a function of pressure. $\rho_0$ is the measured value of the normal state resistivity as $T \rightarrow 0$, revealed by the application of a small magnetic field (see text and Fig.~\ref{fig3} and \ref{fig4}). Bottom right: inelastic part $\Delta\rho_a = \rho_a - \rho_{0}$ of the resistivity at 1 K. The dashed line is a linear fit to all the data points except that at $T_c$~=~0.87~K
}
\label{fig2}
\end{center}
\end{figure}


\begin{figure}[t!]
\begin{center}
\includegraphics[scale=1.1]{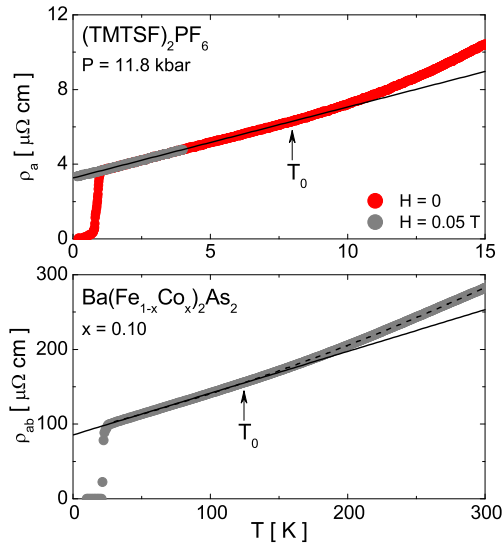}
\caption{(Color online) Top: $a$-axis electrical resistivity $\rho_a(T)$ of \PF~at 11.8 kbar, in $H$ = 0 and 0.05 T. The black line is a linear fit to the data up to $T_0$~=~8~K. Bottom: in-plane ($ab$) electrical resistivity of Ba(Fe$_{1-x}$Co$_x$)$_2$As$_2$ at $x$~=~0.10 (reproduced from \cite{Fang}). The black line is a linear fit from $T_c$ up to $T_0$~=~125~K. The dashed line is a polynomial fit of the form $\rho(T) = \rho_0 + AT + BT^2$ from $T_c$ up to 300 K.
}
\label{fig3}
\end{center}
\end{figure}


In Fig.~\ref{fig2} we show the zero-field electrical resistivity of \PF~for a range of pressures that nearly span the entire superconducting phase, from 8.4 up to 20.8 kbar. At $P$ = 8.4~kbar, upon cooling the resistivity first rises suddenly when \tsdw~is crossed and \PF~enters the SDW state, and then drops sharply at the superconducting $T_c$. Increasing the pressure further completely suppresses the SDW phase, and brings a smooth and monotonic reduction of $T_c$ and of the resistivity. Samples of organic matter are susceptible to forming cracks caused by thermal cycling or pressurization, which renders them useless for absolute measurements. We have measured a number of samples and here we report data for a specimen which showed no sign of cracks, therefore corresponding to the intrinsic evolution with pressure and temperature of the resistivity of \PF. For instance, the resistance of this specimen showed no sudden change during pressurization and it returned to its initial value after each cooling cycle. The weak pressure dependence of the residual resistivity $\rho_0$, shown in Fig.~\ref{fig2}, further confirms the absence of significant cracks.

\begin{figure}[t!]
\begin{center}
\includegraphics[scale=1.2]{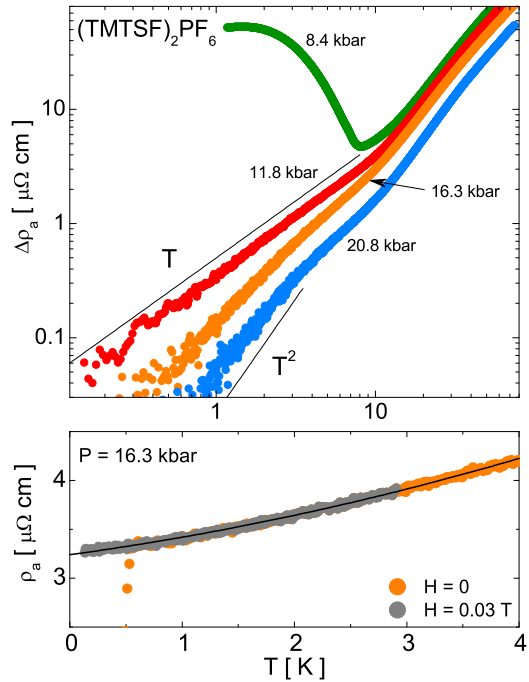}
\caption{(Color online) Top: inelastic part $\Delta\rho_a(T) = \rho_a(T) - \rho_{0}$ of the normal-state $a$-axis electrical resistivity of \PF~at 8.4, 11.8, 16.3, and 20.8~kbar, in a small magnetic field of typically 0.05 T. The lines represent $\Delta\rho(T) \propto T$ and $\Delta\rho(T) \propto T^2$. Bottom: $a$-axis electrical resistivity of \PF~at 16.3 kbar, in $H$=0 and 0.03 T. The black line is a polynomial fit of the form $\rho(T) = \rho_0 + AT + BT^2$ from 0.1 up to 4.0 K.
}
\label{fig4}
\end{center}
\end{figure}

At $P$ = 11.8~kbar, near the critical pressure where \tsdw~$\rightarrow 0$ and $T_c$ is the highest~\cite{Vuletic,Salameh}, the resistivity decreases monotonically with decreasing temperature and displays a strict linear temperature dependence below an upper temperature $T_0$~=~8~K, as seen in Fig.~\ref{fig3}. The application of a small magnetic field of $H$ = 0.05 T, whose sole effect, at all pressures, is to reveal the normal state resistivity below $T_c$ without any magnetoresistance, as seen in Fig.~\ref{fig3} and \ref{fig4}, shows that this pure linear resistivity extends to the lowest measured temperature, thus covering nearly two decades in temperature, from $\sim$~0.1 up to 8.0~K. This finding is further emphasized in Fig.~\ref{fig4}, where only the inelastic part of the normal state resistivity is plotted on a log-log scale.

\begin{figure}[t!]
\begin{center}
\includegraphics[scale=1.1]{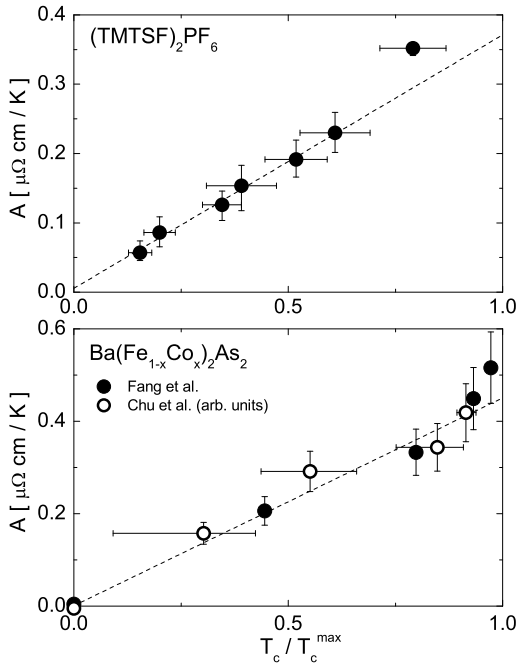}
\caption{Top: Coefficient $A$ of linear resistivity as a function of normalized $T_c$ ($T_c^{\textrm{max}}$ = 1.1 K) for \PF, from a second order polynomial fit over the range 0.1 - 4.0 K to all our resistivity curves at different pressure points between 11.8 and 20.8 kbar. The vertical error bars show the variation of $A$ when the upper limit of the fit is changed by $\pm$~1.0 K. $T_c$ is defined as the midpoint of the transition and the error bars come from the 10~\% and 90~\% points. The dashed line is a linear fit to all the data points except that at $T_c$~=~0.87~K. Bottom: Coefficient $A$ of linear resistivity as a function of normalized $T_c$ ($T_c^{\textrm{max}}$ = 26 K) for Ba(Fe$_{1-x}$Co$_x$)$_2$As$_2$. The data points come from a second order polynomial fit over the range 30 - 300 K (except for $x$~=~0.184 and 0.30, where $T_c$~=~0 and the range is 0 - 300 K) to the resistivity data of Fang \etal~(red dots)~\cite{Fang} and Chu \etal~(blue dots)~\cite{Chu}. For the latter, $A$ is expressed in arbitrary units. The vertical error bars come from an estimated $\pm$~15~\% uncertainty on the geometric factors. The $T_c$ and corresponding error bars are those quoted in~\cite{Fang,Chu}. The dashed line is a guide to the eye.
}
\label{fig5}
\end{center}
\end{figure}

As the pressure is further increased, the low temperature part of the resistivity below 4~K acquires a curvature which, as seen in Fig.~\ref{fig4}, approaches a $T^2$ dependence at the highest measured pressure of 20.8~kbar, close to the pressure where the superconducting $T_c$ vanishes. At intermediate pressures, it is a sum of linear and $T^2$ terms that seems to best describe the low temperature data, as shown in the bottom panel of Fig.~\ref{fig4} where the resistivity at 16.3~kbar is well fitted by a polynomial function of the form $\rho(T) = \rho_0 + AT + BT^2$, from 0.1~K up to 4~K. We note that early data on \PF~at one pressure point \cite{Schulz} also display a non-Fermi-liquid temperature dependence of this kind, although it was not recognized as such at the time. Using this approximate description, we track the evolution of the linear resistivity with pressure, from the critical point for SDW order to that for superconductivity, which reveals our central finding, shown in Fig.~\ref{fig5}: the coefficient $A$ of linear resistivity scales with $T_c$ and vanishes at the point where superconductivity ceases to exist.

We have performed the same set of measurements in a second Bechgaard salt, \ClO, and the very same correlation between $A$ and $T_c$ was observed~\cite{NDL}. Because \ClO~is a high-pressure analogue of \PF, with an ambient-pressure $T_c$ of $\sim$1.2~K, using the same experimental setup we were able to well exceed the point at which $T_c$~=~0, thus confirming that $A$~=~0 when $T_c$~=~0~\cite{NDL}.


We stress that this correlation between resistivity and $T_c$ is not fit-dependent. The purely linear and nearly $T^2$ regimes close to the SDW and superconducting critical points, respectively, are clear from the resistivity curves (Figs.~\ref{fig3} and \ref{fig4}). Describing the evolution from one regime to the other as we did allows us to track each contribution, $AT$ and $BT^2$, as a function of pressure. But a power law of the form $\rho(T) = \rho_0 + AT^{\alpha}$ would also describe the data well, giving an exponent $\alpha$ that grows from 1 to 2 with pressure. An alternative measure of the linear resistivity which involves no fit consists in taking the inelastic part $\Delta\rho_a = \rho_a - \rho_0$ of the resistivity at 1K - where the $T^2$ term is negligible - and plotting that quantity versus $T_c$. As shown in Fig.~\ref{fig2}, the correlation between $\Delta\rho$ at 1K and $T_c$ is the same as that seen between $A$ and $T_c$, \ie, it extrapolates to the origin, showing that it is not the result of a particular fitting procedure.


The observation of a strict linear resistivity as $T \rightarrow 0$ in the Bechgaard salt \PF~on the verge of SDW order is highly reminiscent of the linear resistivity seen in heavy fermion metals at an antiferromagnetic quantum critical point where it is ascribed to fluctuations of the incipient magnetic order \cite{vL,Gil}. The correlation between linear resistivity and $T_c$ established here now shows that scattering and pairing share a common origin, implying that antiferromagnetic spin fluctuations and superconductivity are intimately connected, as discussed in the context of heavy fermion, ruthenate, and cuprate superconductors \cite{Monthoux}.

A weak-coupling solution to the problem of the interplay between antiferromagnetism and superconductivity in the Bechgaard salts has been worked out using the renormalization group approach \cite{Sedeki,Nickel}. The calculated phase diagram captures the essential features of the experimentally-determined phase diagram of \PF~\cite{Sedeki}. The superconducting state below $T_c$ has $d$-wave symmetry \cite{Shinagawa}, with pairing coming from antiferromagnetic spin fluctuations. The normal state above $T_c$ is characterized by the constructive interference of antiferromagnetic and pairing correlations, which \textit{enhances} the amplitude of spin fluctuations \cite{Sedeki,Nickel}. The antiferromagnetic correlation length $\xi(T)$ increases according to $\xi = c(T + \Theta)^{-1/2}$ as $T \rightarrow T_c$, where $\Theta$ is a positive temperature scale \cite{Sedeki}. This correlation length is expected to impart an anomalous temperature dependence to any quantity that depends on spin fluctuations. For instance, it was shown \cite{Sedeki} to account in detail for the deviation from the Fermi-liquid behavior in the NMR relaxation rate measured in the Bechgaard salts \cite{Creuzet,Shinagawa}. Through Umklapp scattering, antiferromagnetic spin fluctuations will also convey an anomalous temperature dependence to the quasi-particle scattering rate $\tau^{-1}$, in addition to the regular Fermi-liquid component which goes as $T^2$. Evaluation of the imaginary part of the one-particle self-energy yields $\tau^{-1} = aT\xi  + bT^2$, where $a$ and $b$ are constants. It is then natural to expect the resistivity to contain a linear term $AT$ (in the limit $T \ll \Theta$), whose magnitude would presumably be correlated with $T_c$, as both scattering and pairing are caused by the same antiferromagnetic correlations. Calculations of the conductivity are needed to see whether the combined effect of pairing and antiferromagnetic correlations conspires to produce the remarkably linear resistivity observed in \PF~on the border of SDW order. 

Comparison with the resistivity data of Fang \etal~\cite{Fang} and Chu \etal~\cite{Chu} on the pnictide superconductor Ba(Fe$_{1-x}$Co$_x$)$_2$As$_2$ suggests that our findings on the Bechgaard salts may be a more general property of metals near a SDW instability. The phase diagram of Ba(Fe$_{1-x}$Co$_x$)$_2$As$_2$ \cite{Chu,Fang}, shown in Fig.~\ref{fig1}, is strikingly similar to that of \PF, with \tsdw~and $T_c$ both enhanced by a factor of about 20, and just above the critical doping where \tsdw~$\rightarrow$~0 (at $x \approx$~0.08), its resistivity is purely linear below $T_0 \approx$~125 K, down to at least $T_c \approx$~25 K (Fig.~\ref{fig3}). We note that the ratio of $T_0$ to the maximum $T_c$ is roughly the same for \PF~and Ba(Fe$_{1-x}$Co$_x$)$_2$As$_2$. Furthermore, in the overdoped regime ($x >$~0.08) and over a large temperature range, from $T_c$ up to 300 K, the resistivity of Ba(Fe$_{1-x}$Co$_x$)$_2$As$_2$ \cite{Fang,Chu} is well described by $\rho(T) = \rho_0 + AT + BT^2$ (see Fig.~\ref{fig3}), with a linear coefficient $A$ that decreases monotonically as $T_c$ drops (Fig.~\ref{fig5}), vanishing at the critical doping $x_c \approx$~0.18 where $T_c \rightarrow$~0. For $x = x_c$ and beyond, $A = 0$ \cite{Fang,Chu}. This reveals a detailed similarity with \PF, which further reinforces the connection between linear resistivity, antiferromagnetic fluctuations and superconductivity described above.

In the cuprates, it has long been known that the low-temperature resistivity of strongly-overdoped non-superconducting samples has the form $\rho(T) = \rho_0 + BT^2$, as in Tl$_2$Ba$_2$CuO$_{6+\delta}$ (Tl-2201) at $p = 0.27$ \cite{Manako} and La$_{2-x}$Sr$_x$CuO$_4$ (LSCO) at $p = 0.33$ \cite{Nakamae}. It was also shown that the evolution of $\rho(T)$ from  $\rho(T) = \rho_0 + AT$ near optimal doping to $\rho(T) = \rho_0 + BT^2$ at high doping is best described by the approximate form  $\rho(T) = \rho_0 + AT + BT^2$ at intermediate doping \cite{Mackenzie,Proust}. The $A$ coefficient thus obtained, when expressed per CuO$_2$ plane, \ie, $A/d$, where $d$ is the average distance between CuO$_2$ planes, was recently found to be universal among hole-doped cuprates and shown to correlate with $T_c$, vanishing at the doping where superconductivity disappears~\cite{NDL}. The same correlation was found in an analysis of low-temperature resistivity measurements in high magnetic fields on overdoped LSCO~\cite{Cooper}. In the context of cuprates, a linear transport scattering rate was explained in terms of antiferromagnetic fluctuations \cite{Moriya}, or as a property of a marginal Fermi liquid \cite{Varma}


In summary, we have observed a linear-$T$ resistivity as $T~\rightarrow$~0 on the border of SDW order in the Bechgaard salt \PF, showing that it is a property of metals close to a magnetic instability which transcends the peculiarities of $f$-electron metals and their Kondo physics or copper oxides and their Mott physics. Away from the SDW phase, the low-temperature resistivity acquires a curvature and eventually becomes quadratic when $T_c~\to~0$. The correlation between non-Fermi-liquid resistivity (linear) and superconducting $T_c$ reveals that anomalous scattering and pairing have a common origin. In \PF, all evidence suggest that both are caused by antiferromagnetic spin fluctuations. The similar phase diagram, detailed temperature dependence of the resistivity and correlation with $T_c$ observed in Ba(Fe$_{1-x}$Co$_x$)$_2$As$_2$ strongly suggest that antiferromagnetic fluctuations play a similar fundamental role in the pnictide superconductors, with temperature scales \tsdw~and $T_c$ twenty times higher. While the situation in cuprates is more complex, in particular because of the ill-understood pseudogap phase, the fact that the same correlation between non-Fermi-liquid resistivity and $T_c$ is observed outside the pseudogap phase in several cuprates~\cite{NDL,Cooper} would seem to also favour, by analogy, the same scenario, at least in the overdoped regime.


We thank H. Shakeripour for assistance with the data analysis. This work was supported by NSERC (Canada), FQRNT (Qu\'ebec), CFI (Canada), a Canada Research Chair (L.T.), the Canadian Institute for Advanced Research, and CNRS (France).


\end{document}